\documentclass[a4paper,12pt]{article}
\usepackage{epsfig}
\newcommand{\be}{\begin{equation}}
\newcommand{\ee}{\end{equation}}
\newcommand{\bea}{\begin{eqnarray}}
\newcommand{\eea}{\end{eqnarray}}
\newcommand{\norsl}{\normalsize\sl}
\newcommand{\norsc}{\normalsize\sc}

\def \psl {p \kern-.45em{/}}
\def \ksl {k \kern-.55em{/}}
\def \ssl {s \kern-.45em{/}}

\def\ol#1{\overline{#1}}

\def\nn{\nonumber}

\begin{document}

\begin{titlepage}

\title{Zee model and phenomenology of lepton sector }

\author{
\norsc  Eiichi MITSUDA and  Ken SASAKI\thanks{e-mail address:
sasaki@phys.ynu.ac.jp} \\
\norsl  Dept. of Physics,  Faculty of Engineering, Yokohama National
University \\
\norsl  Yokohama 240-8501, JAPAN \\
}

\date{}
\maketitle

\date{}
\maketitle

\begin{abstract}
{\normalsize The virtual effects of the Zee charged scalar boson on  
the lepton-family-number (LFN) violating processes are studied. We obtain 
the constraints on the individual Yukawa coupling constants of the Zee boson to
leptons. Using these constraints, we predict the upper bounds on the 
muonium-antimuonium conversion probability, the branching fractions of 
the LFN violating decays such as  
$\tau\rightarrow e\gamma$, $\tau\rightarrow \mu\gamma$, 
$\tau^-\rightarrow \mu^+e^-e^-$ and $\tau\rightarrow 
e^+\mu^-\mu^-$. 
The contribution of the Zee boson to the muon anomalous magnetic moment 
is also considered.
}
\end{abstract}

\begin{picture}(5,2)(-290,-500)
\put(2.3,-120){YNU-HEPTh-01-102}
\put(2.3,-135){March 2001}
\end{picture}

\thispagestyle{empty}
\end{titlepage}
\setcounter{page}{1}
\baselineskip 18pt

Accumulated data on atmospheric, solar and accelerator neutrino 
experiments indicate that neutrinos have small but finite masses and 
mix among flavors \cite{ICHEP}. Existence of massive neutrinos, even though they are 
very tiny, necessitates the extension of the standard model (SM) for electroweak
interactions. To explain the smallness of neutrino masses, there are two well-known 
mechanisms. One is the see-saw mechanism which introduces heavy right-handed 
Majorana neutrinos \cite{See-saw}. The other is the radiative mechanism in which  
neutrino masses are generated  by radiative corrections and the left-handed
neutrinos acquire the  Majorana masses.  The simplest model for the latter scenario was
presented by  Zee \cite{ZeeModel}. In this model, the ordinary left-handed
neutrinos are employed  while a new charged scalar field $h$, being a ${\rm SU(2)}_L$
singlet, and  two doublets of Higgs bosons  $\Phi_1$, $\Phi_2$ are introduced. 
The neutrino mass matrix in the Zee model, generated  at
one-loop level, shows a very distinctive pattern \cite{Wol,Pet}.  Recently, in 
connection with  neutrino oscillations, the neutrino mass and mixing matrices 
of the Zee model have been extensively studied \cite{SmiTani,JMST}. Especially, it has
been recognized \cite{JMST} that the Zee model yields a solution for the bi-maximal
neutrino mixing both in atmospheric and solar neutrino oscillations \cite{bimax}. 

The charged scalar boson $h$ in the Zee model carries 
lepton number $L(= L_e+L_{\mu}+L_{\tau})=2$, and its Yukawa couplings to leptons 
violate the lepton-family-number (LFN) conservation. Thus the Zee scalar boson $h$ not
only  plays a crucial role for neutrino mass generation, but also 
it induces other interesting LFN violating weak processes. 
In fact, these processes have been studied before in the Zee
model and some constraints on the Yukawa couplings between the Zee boson $h$ and
leptons  have been obtained
\cite{ZeeModel}-\cite{SmiTani},\cite{LeoTamVer}-\cite{McNg}.  

In this paper we further study the virtual effects of the Zee boson on
the LFN violating weak processes. Although some information on the parameters 
in the Zee model has been reported from the analysis of  neutrino oscillations 
within the framework of the Zee mass matrix \cite{JMST}, we refrain from its use 
here. Instead, we employ  
the experimental upper bounds on the muon decay rate, the $g_{\mu}/g_e$ ratio, 
and the $\mu \rightarrow e\gamma$ branching fraction, 
and show that we can obtain the constraints on
the  individual ratios, $\frac{\vert f_{e\mu}\vert^2}{{\overline M_1}^2}$, 
$\frac{\vert f_{e\tau}\vert^2}{{\overline M_1}^2}$, and 
$\frac{\vert f_{\mu\tau}\vert^2}{{\overline M_1}^2}$, 
where $f_{ij}$'s are the Yukawa coupling constants of the Zee boson $h$ to leptons and 
${\overline M_1}$ is the ``Zee boson" mass (see Eq.(\ref{M1line}) below for the
definition).  With these constraints at hand, we analyze other LFN violating processes 
such as the muonium-antimuonium conversion, 
$\tau\rightarrow e\gamma$, $\tau\rightarrow \mu\gamma$, 
$\tau^-\rightarrow \mu^+e^-e^-$ and $\tau\rightarrow 
e^+\mu^-\mu^-$ decays. Finally we consider the contribution of the Zee boson to the 
muon anomalous magnetic moment.

\bigskip

In the Zee model, the following Lagrangian is added to the SM:
\bea
{\cal L}_{\rm Zee}&=&\sum_{i,j=e,\mu,\tau}f_{ij}\psi_{iL}^T C
              (i\sigma_2) \psi_{jL} h^+ +
        \mu \Phi^T_1 (i\sigma_2) \Phi_2 h^- + h.c.   \nn  \\
&=&2f_{e\mu}\Bigl[ \nu_{eL}^TC \mu_L 
  - e_{L}^TC\nu_{\mu L} \Bigr] h^+ + 
2f_{e\tau}\Bigl[ \nu_{eL}^TC\tau_L 
  - e_{L}^T C \nu_{\tau L}  \Bigr] h^+ \nonumber  \\
&+&2f_{\mu\tau}\Bigl[ \nu_{\mu L}^TC \tau_L 
  - \mu_{L}^T  C\nu_{\tau L} \Bigr] h^+ +
\mu\Bigl(\Phi^+_1 \Phi^0_2 - \Phi^0_1 \Phi^+_2  \Bigr)h^-  +h.c.~,
\label{LZEE}
\eea
where $\psi_{iL}=(\nu_i, l_i)^T_L$ (with $i$ a family index) is an usual
left-handed lepton  doublet and $h^{\pm}$ is the ${\rm SU(2)}_L$ singlet Zee scalar
boson. Two Higgs  doublets, $\Phi_j=(\Phi^+_j,\Phi^0_j)^T, j=1,2$,  are introduced and
we assume that only $\Phi_1$ couples to leptons. 
Since $f_{ij}$ is antisymmetric under the interchange of lepton family 
indices $i$ and $j$, the Yukawa couplings of $h$ to leptons 
violate the lepton-{\it family}-number-conservation.
After the neutral components having
acquired the vacuum expectation values, $\langle\Phi^0_j
\rangle=v_j/{\sqrt 2}$,~  the charged Higgs boson (which is orthogonal to the would-be
Goldstone boson) is  expressed as
\be
\Phi^+={\rm cos}\beta~ \Phi_1^+-{\rm sin}\beta~ \Phi_2^+  ~,
\ee
where ${\rm tan}\beta\equiv \frac{v_1}{v_2}$.
This charged Higgs boson $\Phi^+$ mixes with the Zee boson $h^+$ due
to the $\Phi_1$-$\Phi_2$-$h$ interaction given in Eq.(\ref{LZEE}). 
After the charged scalar mass matrix being diagonalized, $h^+$ and $\Phi^+$ are
expressed in terms of  the physical charged scalar bosons $H_1^+$ and $H_2^+$ with mass
eigenvalues $M_1^2$ and $M_2^2$, respectively,  as~\cite{Pet}
\be
\pmatrix{h^+\cr \Phi^+\cr}=\pmatrix{{\rm cos}\phi&{\rm sin}\phi \cr
-{\rm sin}\phi&{\rm cos}\phi \cr}\pmatrix{H_1^+\cr H_2^+\cr}~,
\ee
with
\be
  {\rm sin}2\phi=\frac{2{\sqrt 2}\mu M_W}{g(M_1^2-M_2^2)} ~,
\ee
where $M_W=\frac{g}{2}\sqrt {v_1^2+v_2^2}$, the mass of $W$ gauge boson,  and $g$ is
the $SU(2)_L$ gauge coupling constant. 

Since the physical charged $H_1^{\pm}$ and $H_2^{\pm}$ bosons  are a linear 
combination of $h^{\pm}$ and $\Phi^{\pm}$, the interaction of $H_1$ and $H_2$ with
leptons  is made up of two parts, one being  LFN conserving and 
the other  LFN changing, and it takes the following form:
\bea
{\cal L}^{{\rm lepton}-H_i}&=&\frac{g{\rm cot}\beta}{{\sqrt 2}M_W}
 \Bigl(\sum_{i=e,\mu,\tau} m_i \ol{\nu}_{i L}l_{iR}\Bigr)
\Bigl(-{\rm sin}\phi H_1^+ + {\rm cos}\phi H_2^+    \Bigr) \nonumber \\
&&+\biggl\{ 2f_{e\mu}\Bigl[ \nu_{eL}^TC \mu_L  - e_{L}^TC\nu_{\mu L} \Bigr]  + 
2f_{e\tau}\Bigl[ \nu_{eL}^TC\tau_L - e_{L}^T C \nu_{\tau L}  \Bigr] \nonumber  \\
&&\quad +2f_{\mu\tau}\Bigl[ \nu_{\mu L}^TC \tau_L  - \mu_{L}^T  C\nu_{\tau L} \Bigr] 
\biggr\}\Bigl({\rm cos}\phi H_1^+ + {\rm sin}\phi H_2^+    \Bigr) +h.c.~.
\eea
The terms in the first line are LFN conserving and each term is 
proportional to a lepton mass $m_i$, and the rest are 
LFN changing terms stemming from the $h^{\pm}$-lepton interactions.

\bigskip

The effective four-fermion Lagrangian induced by $H_1$ and $H_2$ exchange gives 
observable contributions to weak processes. The dominant terms for the decays 
$\mu \rightarrow e{\overline \nu}\nu$, $\tau \rightarrow e{\overline \nu}\nu$, 
and $\tau \rightarrow \mu{\overline \nu}\nu$, for example, are given 
by \cite{SmiTani,McNg} 
\bea
-{\cal L}_{\rm eff}&=&\frac{4 G_F}{\sqrt{2}}\biggl\{
\Bigl(1+\frac{\vert f_{e\mu}\vert^2}{\sqrt{2}G_F{\overline M_1}^2}    \Bigr)
\Bigl[{\overline e_L}\gamma_{\lambda}\nu_{eL}\Bigr]
\Bigl[{\overline \nu_{\mu L}}\gamma^{\lambda}\mu_{L}\Bigr]\nonumber  \\
&&\qquad +\Bigl(1+\frac{\vert f_{e\tau}\vert^2}{\sqrt{2}G_F{\overline M_1}^2}    \Bigr)
\Bigl[{\overline e_L}\gamma_{\lambda}\nu_{eL}\Bigr]
\Bigl[{\overline \nu_{\tau L}}\gamma^{\lambda}\tau_{L}\Bigr]\nonumber  \\
&&\qquad +\Bigl(1+\frac{\vert f_{\mu\tau}\vert^2}
{\sqrt{2}G_F{\overline M_1}^2}    \Bigr)
\Bigl[{\overline \mu_L}\gamma_{\lambda}\nu_{\mu L}\Bigr]
\Bigl[{\overline \nu_{\tau L}}\gamma^{\lambda}\tau_{L}\Bigr] \biggr\}~,
\label{Decay}
\eea
where
\be
\frac{1}{{\overline M_1}^2}=  \frac{{\rm cos}^2\phi}{M_1^2}+ \frac{{\rm
sin}^2\phi}{M_2^2}~.  
\label{M1line} 
\ee

The constraint on $\vert f_{e\mu}\vert^2/{\overline M_1}^2$ has been obtained 
from the study of the muon decay rate. Smirnov and Tanimoto \cite{SmiTani} got 
$\vert f_{e\mu}\vert^2/{\overline M_1}^2<7\times10^{-4}G_F$ by assuming that the 
effect of the new bosons on the muon decay rate is smaller than $0.1{\%}$. On the
other hand, McLaughlin and Ng \cite{McNg} obtained 
\be
\frac{\vert f_{e\mu}\vert^2}{{\overline M_1}^2}<3\times10^{-3}G_F~,
\label{ConstFeu}
\ee
by demanding that the corrections be no bigger than 
the error of the SM Fermi constant. In this paper we take the latter one, 
a rather conservative constraint for $\vert f_{e\mu}\vert^2/{\overline M_1}^2$. 
Authors of Ref.\cite{McNg} also pointed out that an information on the difference 
$(\vert f_{e\tau}\vert^2-\vert f_{\mu\tau}\vert^2)/{\overline M_1}^2$ can be obtained 
from the ratio of $\tau$ decay rates 
$\Gamma(\tau \rightarrow \mu{\overline \nu}\nu)/
\Gamma(\tau \rightarrow e{\overline \nu}\nu)$.  Quite recently  new results 
have been reported on the branching fractions of $\tau$ into leptons
\cite{L3} and the ratio of the charged current coupling constants of the muon and
electron is determined to be $g_{\mu}/g_e=1.0007\pm0.0051$. The Zee model predicts 
(see Eq.(\ref{Decay})),
\be
\frac{g_\mu}{g_e}=1+\frac{1}{\sqrt{2}G_F}
\frac{\vert f_{\mu\tau}\vert^2-\vert f_{e\tau}\vert^2}{{\overline M_1}^2}~.
\ee
Taking the bound $\vert \frac{g_\mu}{g_e}-1   \vert <0.006$, we obtain 
\be
\biggl|\frac{\vert f_{\mu\tau}\vert^2}{{\overline M_1}^2}-\frac{\vert
f_{e\tau}\vert^2}{{\overline M_1}^2}\biggr| <8.5\times 10^{-3}G_F~.
\label{GmuoverGe}
\ee
This  gives only the upper bound on the difference. However, 
if we combine this constraint with the  one from the  
$\mu \rightarrow e\gamma$ decay, we obtain the individual bounds on 
 $\vert f_{\mu\tau}\vert^2/{\overline M_1}^2$ and 
$\vert f_{e\tau}\vert^2/{\overline M_1}^2$.
In the Zee model, the branching fraction for $\mu \rightarrow e\gamma$ is given 
by~\cite{Pet}
\be
B(\mu \rightarrow e\gamma)=\frac{\alpha}{48\pi}
\Bigl(\frac{\vert f_{\mu\tau}f_{e\tau}\vert}{G_F{\overline M_1}^2}   \Bigr)^2~.
\label{mueg}
\ee
The present experimental upper pound, 
$B(\mu \rightarrow e\gamma)<1.2\times 10^{-11}$~\cite{Particle}, leads to 
\be
\frac{\vert f_{\mu\tau}f_{e\tau}\vert}{{\overline M_1}^2} <2.8\times 10^{-4}G_F~.
\label{MuEGamma}
\ee 
Then we find that the above constraint, combined with Eq.(\ref{GmuoverGe}), gives the
bounds
\be
\frac{\vert f_{\mu\tau}\vert^2}{{\overline M_1}^2},\ \  \frac{\vert
f_{e\tau}\vert^2}{{\overline M_1}^2} <8.5\times 10^{-3}G_F~.
\label{ConstFetau}
\ee
The fact that $\vert f_{\mu\tau}\vert^2/{\overline M_1}^2$ and 
$\vert f_{e\tau}\vert^2/{\overline M_1}^2$  individually get almost the same 
bound as their difference is due to the very stringent constraint of 
Eq.(\ref{MuEGamma}). Of course, if we get more precise determination of 
the  $g_{\mu}/g_e$ ratio, we can set more stringent bounds on 
$\vert f_{\mu\tau}\vert^2/{\overline M_1}^2$ and 
$\vert f_{e\tau}\vert^2/{\overline M_1}^2$.
 
With these bounds on the Yukawa coupling constants $f_{ij}$, 
Eqs.(\ref{ConstFeu}), (\ref{MuEGamma}), (\ref{ConstFetau}),  
we now discuss the weak processes involving charged $H_1$ and $H_2$ 
bosons. In particular we are interested in the LFN violating processes.

\bigskip

{\it The muonium-antimuonium conversion}.  In the Zee model, the mixing of muonium $M(\mu^+e^-)$ and antimuonium  
${\overline M}(\mu^-e^+)$ arises, at one-loop-level, only through one diagram, 
the one with  two $H_i$ exchange illustrated in Fig.1(a). 
Using a formula for the loop integral
\bea
i\int\frac{d^4k}{(2\pi)^4}\frac{1}{k^2-M^2_i}\frac{1}{k^2-M^2_j}
\Bigl[\frac{1}{\ksl}  \Bigr]\Bigl[\frac{1}{\ksl}  \Bigr]
&=&\frac{1}{64\pi^2}\frac{1}{M^2_i-M^2_j}{\rm ln}\frac{M^2_i}{M^2_j}
\Bigl[\gamma_{\lambda} \Bigr]\Bigl[\gamma^{\lambda}   \Bigr]~\quad (i\ne j)~,
\nonumber\\
&=&\frac{1}{64\pi^2}\frac{1}{M^2_i}\Bigl[\gamma_{\lambda}
\Bigr]\Bigl[\gamma^{\lambda}   \Bigr]~\qquad (i= j)~, \nonumber
\eea
where $[1/\ksl]$ is coming from the internal neutrino propagator, we obtain 
the following effective Lagrangian for the
$M$-${\overline M}$ conversion:
\be
-{\cal L}_{\rm eff}^{M{\overline M}}=\frac{G_{M{\overline M}}}{\sqrt 2}
\Bigl[{\overline \mu}\gamma_{\lambda}(1-\gamma_5)e  \Bigr]
\Bigl[{\overline \mu}\gamma^{\lambda}(1-\gamma_5)e  \Bigr]~,
\ee
with
\be
\frac{G_{M{\overline M}}}{\sqrt 2}=\frac{\vert f_{e\tau} f_{\mu\tau}\vert^2}{16\pi^2}
\frac{1}{{\widetilde M}^2}~,
\label{GMM}
\ee
where
\be
\frac{1}{{\widetilde M}^2}= \frac{{\rm cos}^4\phi}{M_1^2}
+\frac{2 {\rm cos}^2\phi~ {\rm sin}^2\phi}{M_1^2-M_2^2} {\rm ln} 
\frac{M_1^2}{M_2^2} + \frac{{\rm sin}^4\phi}{M_2^2}~. 
\label{Mtilde}
\ee
Note that ${\cal L}_{\rm eff}^{M{\overline M}}$ is in the 
$(V-A)\times(V-A)$ form\footnote{The $M$-${\overline M}$ conversion experiments have
been performed under the influence of external magnetic field. 
The $M$-${\overline M}$ conversion probability reduces with the applied magnetic  
field, and the reduction depends on the effective Hamiltonian being in the form of
$(V\mp A)\times(V\mp A)$, or $(V\mp A)\times(V\pm A)$ or other interaction types
\cite{Fujii}}.  The integrated
probability that the muonium
$M(\mu^+e^-)$ decays as $\mu^-$  rather than $\mu^+$ is given by~\cite{FeinWein}
\be
P({\overline M})=64G_{M{\overline M}}^2/\pi^2 a_B^6 \lambda^2~,
\ee
where $a_B$ is the Bohr radius $(m_e\alpha)^{-1}$ and 
$\lambda=G_F^2m_{\mu}^5/192\pi^3$
is the muon decay rate. 
Using the bound (\ref{MuEGamma}) and the relation
\be
\frac{1}{{\widetilde M}^2}\le\frac{1}{{\overline M_1}^2}\qquad 
{\rm for\ \ arbitrary~\ }\phi~,
\ee
(the equality holds when  $M_1^2=M_2^2$), we obtain 
\be
P({\overline M})< 9.5\times 10^{-24}~\biggl[\frac{{\overline M_1}^2}{M_W^2} 
\biggr]^2~.
\ee
For ${\overline M_1}=800$ GeV, we get $P({\overline M})<10^{-19}$. 
The present experimental upper limit~\cite{Muonium} is 
$P(\ol{M})< 2.4\times 10^{-10}$ (90\% C.L.). 

In fact, the muonium-antimuonium oscillation was studied before in the Zee model 
and the same box diagram in Fig.1(a) was analyzed \cite{LeoTamVer,Ver}. But our result
on the  effective Lagrangian for the muonium-antimuonium conversion disagrees with the
one given in Ref.\cite{LeoTamVer,Ver}, in respect to its chiral structure and
magnitude.  Comparing with  Eqs.(4.116-117) of Ref.\cite{Ver}, we find that our result
on 
$\frac{G_{M{\overline M}}}{\sqrt 2}$ in Eq.(\ref{GMM}) is without a
suppression factor $(m_{\mu}/{\mu_S})^2$, where $\mu_S$ is the charged Zee
boson mass introduced there, expected to be the same order of 
magnitude as our ${\overline M_1}$.

\bigskip

{\it The $\tau\rightarrow e\gamma$ and $\tau\rightarrow \mu\gamma$ decays}.
With replacement of 
$f_{\mu\tau}f_{e\tau}$ in Eq.(\ref{mueg}) with appropriate $f_{ij}$, 
the branching fractions for  both decays are given by 
\bea
B(\tau \rightarrow e\gamma)&=&\frac{\alpha}{48\pi}
\biggl[\frac{\vert f_{e\mu}f_{\mu\tau}\vert}{G_F{\overline M_1}^2}   \biggr]^2
B(\tau \rightarrow e{\overline \nu_{e}}\nu_{\tau}) ~,\\
B(\tau \rightarrow \mu\gamma)&=&\frac{\alpha}{48\pi}
\biggl[\frac{\vert f_{e\mu}f_{e\tau}\vert}{G_F{\overline M_1}^2}   \biggr]^2
B(\tau \rightarrow e{\overline \nu_{e}}\nu_{\tau})~.
\eea
The constraints (\ref{ConstFeu}), (\ref{ConstFetau}) and 
$B(\tau \rightarrow e{\overline
\nu_{e}}\nu_{\tau})_{\rm exp}=(17.83\pm 0.06){\%}$
\cite{Particle} lead to
\be
B(\tau \rightarrow e\gamma)~,\ \  B(\tau \rightarrow \mu\gamma)<2.0\times 10^{-10}~,
\ee
both being far below the present experimental upper bounds 
($\sim 10^{-6}$) \cite{Particle} . 

\bigskip

{\it The $\tau^-\rightarrow \mu^+e^-e^-$ and $\tau\rightarrow 
e^+\mu^-\mu^-$ decays}. The decay $\tau^-\rightarrow \mu^+e^-e^-$ arises,  
at one-loop level, from  
the diagram with two $H_i$ boson exchange depicted in Fig.1(b). The loop integral 
gives 
\be
{\cal L}^{\tau \rightarrow \mu ee}=
\frac{C^{\tau \rightarrow \mu ee}}{\sqrt 2}
\Bigl[{\overline e}\gamma_{\lambda}(1-\gamma_5)\tau  \Bigr]
\Bigl[{\overline e}\gamma^{\lambda}(1-\gamma_5)\mu \Bigr]~,
\ee
with 
\be
\frac{C^{\tau \rightarrow \mu ee}}{\sqrt 2}=
\frac{{f^*_{\mu\tau}}^2 f_{e\tau}f_{e\mu}}{16\pi^2}
\frac{1}{{\widetilde M}^2}~.
\label{TauMuEE}
\ee
Keeping in mind the antisymmetrization of
the final two identical electrons and a symmetry factor $\frac{1}{2}$, and 
neglecting the muon and electron masses, we obtain,  
 $\Gamma(\tau^-\rightarrow \mu^+e^-e^-)=
\vert C^{\tau \rightarrow \mu ee} \vert^2  m_\tau^5/96\pi^3$, 
for the decay rate.
Then we get the upper limit on the branching fraction of 
the $\tau^-\rightarrow \mu^+e^-e^-$ decay,
\be
B(\tau^-\rightarrow \mu^+e^-e^-)<2\times 10^{-18}~\biggl[\frac{{\overline
M_1}^2}{M_W^2}  \biggr]^2~.
\ee

Similarly, the contribution to the decay $\tau\rightarrow  e^+\mu^-\mu^-$ 
comes from the one-loop diagram in Fig.1(c). Now the coefficient
$C^{\tau \rightarrow \mu ee}/\sqrt 2$ in Eq.(\ref{TauMuEE}) is replaced with
$C^{\tau \rightarrow  e\mu\mu}/\sqrt 2=
{f^*_{e\tau}}^2 f_{\mu\tau}f_{e\mu}/16\pi^2 {\widetilde M}^2$, and we obtain the 
same upper limit on $B(\tau^-\rightarrow e^+\mu^-\mu^-)$ 
as the $\tau^-\rightarrow \mu^+e^-e^-$ decay. 
For ${\overline M_1}=800$ GeV, we get the bound $2\times 10^{-14}$ for the 
branching fractions of both decays. 
The present experimental upper limits on both decays are 
$B_{\rm exp}< 1.5\times 10^{-6}$ (90\% C.L.)~\cite{Particle}. 

\bigskip

{\it The} $\mu \rightarrow 3e$~{\it decay}.
This  process is possible in all models which allow the $\mu \rightarrow e\gamma$. 
The photon is now virtual and decays into a $e^+e^-$ pair. Contributions of
the $Z$ boson-exchange graphs instead of photon are negligible.
Thus the branching 
ratio has been estimated to be $B(\mu \rightarrow 3e)\approx (\alpha/\pi) 
B(\mu \rightarrow e\gamma)$~\cite{Pet}. 
At one-loop level in the Zee model, there is another diagram 
contributing to the decay $\mu \rightarrow 3e$, a box diagram
resembling the ones in Fig.1 with two $H_i$ 
exchange.\footnote{The $\mu \rightarrow 3e$ decay was studied before in the Zee model 
\cite{LeoTamVer,Ver}, but our result on
$C^{\mu \rightarrow 3e}_{\rm box}$  disagree with the one in
Refs.\cite{LeoTamVer,Ver}, which has further a suppression factor
$(m_{\mu}/{\mu_S})^2$.}  The diagram gives 
\be
-{\cal L}^{\mu \rightarrow 3e}_{\rm box}=
\frac{C^{\mu \rightarrow 3e}_{\rm box}}{\sqrt 2}
\Bigl[{\overline e}\gamma_{\lambda}(1-\gamma_5)\mu  \Bigr]
\Bigl[{\overline e}\gamma^{\lambda}(1-\gamma_5)e  \Bigr]~,
\ee
with 
\be
\frac{C^{\mu \rightarrow 3e}_{\rm box}}{\sqrt 2}=
\frac{f_{e\tau} f^*_{\mu\tau}}{16\pi^2}
\Bigl[\vert f_{e\mu}\vert^2+\vert f_{e\tau}\vert^2\Bigr]\frac{1}{{\widetilde M}^2}~.
\ee
Assuming that this box diagram contribution adds to the 
decay rate incoherently, we obtain the ratio
\be
\frac{\Gamma(\mu \rightarrow 3e)_{\rm box}}{\Gamma(\mu \rightarrow e\gamma)~
\frac{\alpha}{\pi}}
=12\biggl[\frac{\vert f_{e\mu}\vert^2+\vert f_{e\tau}\vert^2}
{4\pi\alpha}    \biggr]^2\Bigl(\frac{{\overline M_1}}
{{\widetilde M}}   \Bigr)^4~.
\ee
For ${\overline M_1}=800$ GeV, the constraints (\ref{ConstFeu}) and (\ref{ConstFetau})
give $\vert f_{e\mu}\vert^2 <0.02$ and $\vert f_{e\tau}\vert^2<0.06$, 
respectively. Thus, in case 
${\overline M_1}\sim {\widetilde M}$, there remains a possibility that 
this box diagram contribution to the decay rate $\mu \rightarrow 3e$
becomes comparable to the one from the photon exchange graph.

The same remarks can be made for such $\tau$ decays as
$\tau \rightarrow 3e$,  $\tau^- \rightarrow e^-\mu^+\mu^-$, 
$\tau \rightarrow 3\mu$, and $\tau^- \rightarrow \mu^-e^+e^-$. 
The branching fractions for the first two decays are expected 
to be of the order of $(\alpha/\pi) B(\tau \rightarrow e\gamma)$ and 
those for the latter two are $(\alpha/\pi) B(\tau \rightarrow \mu\gamma)$. 
In the Zee model, the two $H_i$ exchange diagrams also contribute to these 
LFN violating decays, and their contributions may possibly be comparable to those 
from the photon exchange diagrams.

\bigskip

{\it The muon anomalous magnetic moment}.  With  improvement in the measurment, 
the anomalous magnetic moment of the muon, $a_{\mu}=(g-2)/2$, has become to
provide an excellent laboratory for testing various electroweak gauge models. 
Quite recently, the new experimental result on the positive muon anomalous magnetic
moment was reported \cite{BNL} and it shows a clear difference between the weighted
mean of the experimental results and the SM prediction:
\be
a_{\mu}({\rm exp})-a_{\mu}({\rm SM})=43(16)\times 10^{-10}~.
\label{ExpSM}
\ee
In the Zee model, two new charged scalar mesons $H_1$ and $H_2$ appear. 
We now discuss their contribution to the muon anomalous magnetic moment.
There are two relevant diagrams in one loop, which are depicted in Fig.2.
The arrows on the fermion lines indicate the flow of lepton number. 
The diagram in Fig.2(a) is LFN conserving and thus  the factor 
$(m_{\mu}/M_W)^2$ appears, while  
the one in Fig.2(b) is LFN changing and its contribution is 
proportional to  
$\vert f_{e\mu}\vert^2$ ($\vert f_{\mu\tau}\vert^2$). The calculation is
straightforward
\cite{L-MWY}, and we find 
\bea
a_{\mu}^{(a)}&=&\frac{m_{\mu}^4}{24\pi^2}
\frac{g^2 {\rm cot}^2\beta}{8M_W^2}
\frac{1}{{\overline M_2}^2}~, \\
 a_{\mu}^{(b)}&=&\frac{m_{\mu}^2}{24\pi^2}\frac{\vert f_{e\mu}\vert^2+\vert
f_{\mu\tau}\vert^2}{{\overline M_1}^2}~,   
\eea
where 
\be
\frac{1}{{\overline M_2}^2}=  \frac{{\rm
sin}^2\phi}{M_1^2}+ \frac{{\rm cos}^2\phi}{M_2^2}~.  
\label{M2line} 
\ee
The LFN conserving contribution is rewritten as
$a_{\mu}^{(a)}=9\times10^{-16}{\rm cot}^2\beta(M_W^2/{\overline M_2}^2)$. 
So even when we take generous values for ${\rm cot}\beta$ and ${\overline M_2}$, say, 
${\rm cot}\beta\sim 50$ and ${\overline M_2}\sim 200$ GeV,  we get $a_{\mu}^{(a)}\sim
4\times 10^{-13}$. For the LFN violating part, 
the bounds (\ref{ConstFeu}) and (\ref{ConstFetau}) give 
$a_{\mu}^{(b)}<6\times10^{-12}$. Thus we observe that the contribution of 
the charged scalar bosons $H_1$ and $H_2$ to the muon 
anomalous magnetic moment is not sufficient fo fill the difference
(\ref{ExpSM}). In addition to these charged scalar bosons, 
there appear three neutral Higgs bosons  in the Zee model. But 
their contributions to the muon anomalous magnetic moment are 
proportinal to  $(m_{\mu}^2/M_W M_H)^2$, where $M_H$ is a
relevant Higgs mass, and, therefore, they are negligible. So we conclude that  the Zee
model in its original form cannot fill the present gap (\ref{ExpSM}) between the
experiment and  the SM prediction.

\bigskip

The Zee model is one of the promising candidates which may explain the 
phenomena of atmospheric and solar neutrino oscillations. Also it predicts many
interesting  LFN violating weak processes. Gaining the current constraints on the 
individual Yukawa coupling constants, $\vert f_{e\mu}\vert^2/{\overline M_1}^2$, 
$\vert f_{e\tau}\vert^2/{\overline M_1}^2$, and 
$\vert f_{\mu\tau}\vert^2/{\overline M_1}^2$, in the Zee model, we have studied 
the virtual effects of the Zee boson on the phenomenology of lepton sector. 
The predicted effects on the LFN violating 
weak processes are found to be small and  far below
the  present experimental upper limits. Also it is found that the contribution of the
new bosons  in the Zee model to the muon
anomalous magnetic  moment is  too small to fill the present difference
between the experiment and  the SM prediction. If this gap persists, we are compelled
to abandan the Zee model  or to pursue the extensions  of its original 
form\footnote{An extension of the Zee model which includes singlet neutrinos has been
considered in Ref.\cite{McNg}.}. Embedding the original Zee model into  supersymmetry
might be one of the most interesting extensions
\cite{SusyZee}.

\bigskip

After submitting the paper, our attension has been called to the following 
two papers, Ref.\cite{BiSanta} and Ref.\cite{DicusHeNg}. 
In the first reference.\cite{BiSanta}, an effective field theory has been built by 
integrating  out the heavy scalar in the Zee model. 
The effective lagrangians obtained for the four-fermi interactions 
are consistent with our results. In the second reference \cite{DicusHeNg}, it was 
pointed out also that the minimal Zee model cannot resolve the BNL $g-2$ anomaly, 
using the neutrino oscillation data, and the extensions of the Zee model 
have been considered to explain the mass patterns of the neutrinos and leptons and
the BNL $g-2$ anomaly.

\vspace{1.5cm}

One of the authors (K.S) would like to thank M. Bando, N. Haba, T. Kugo, and M.
Tanimoto for helpful  discussions. This work is supported by the 
Grant-in-Aid for Scientific Research on Priority Areas (A), Ministry of Education,
Japan (No. 12047212).


\newpage
\vspace{0.5cm}
\leftline{\large\bf Figure Caption}
\vspace{0.5cm}

\noindent
Figure 1

\noindent
The two-$H_i$-exchange box diagrams relevant for (a) the muonium-antimuonium
conversion,  (b) the $\tau \rightarrow
\mu^+e^- e^-$  decay, and (c) the $\tau \rightarrow
e^+\mu^-\mu^-$  decay. The arrows show the flow of lepton number.

\bigskip

\bigskip

\noindent
Figure 2

\noindent
The charged $H_i$ scalar contributions to $a_{\mu}$: (a) the 
LFN conserving diagram and (b) the 
LFN changing diagram. The arrows show the flow of 
lepton number.

\clearpage
\thispagestyle{empty}

\begin{figure}
\begin{center}
\label{fig:test}
\psfig{file=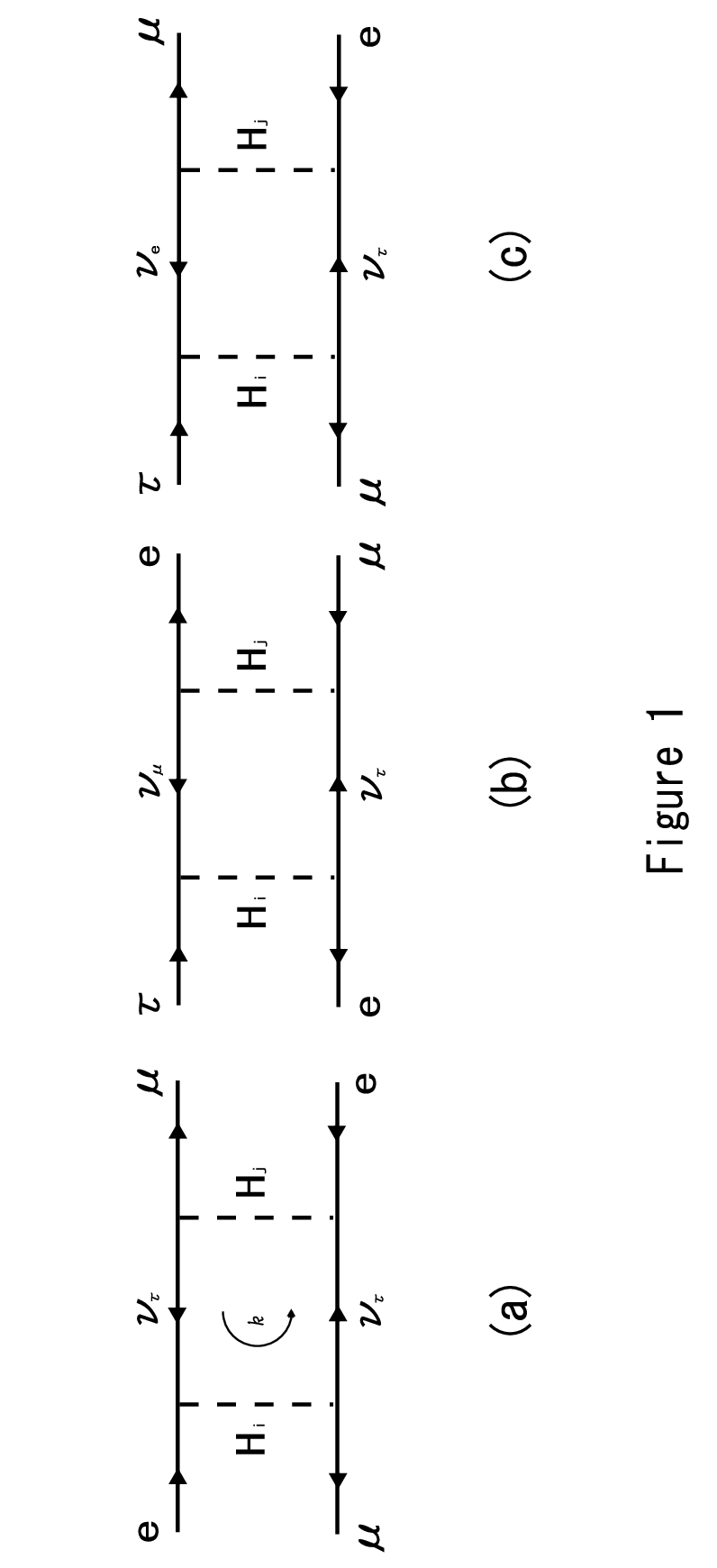,width=9cm}
\end{center}
\end{figure}

\clearpage
\thispagestyle{empty}

\begin{figure}
\begin{center}
\label{fig:test}
\psfig{file=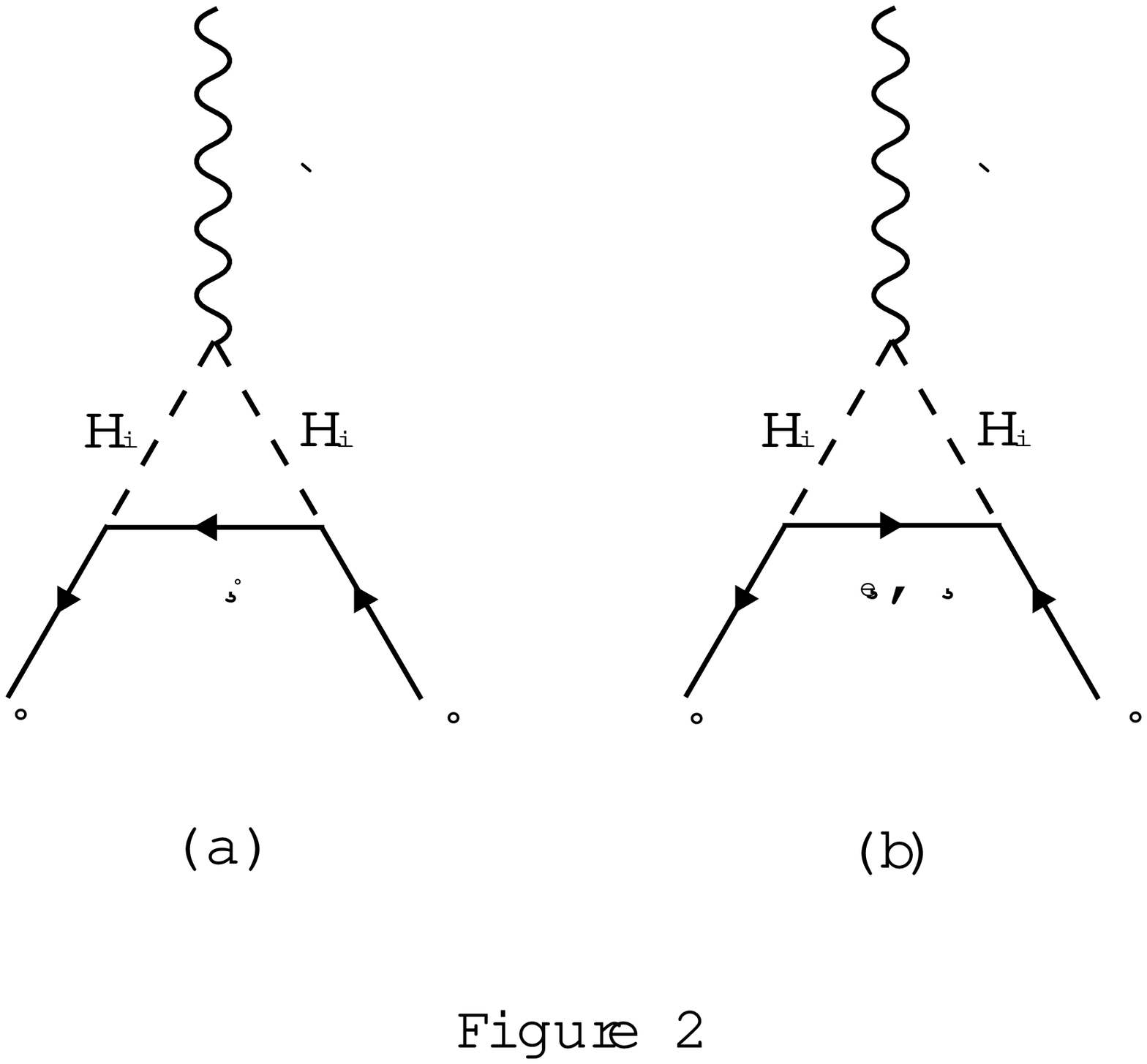,width=5in}
\end{center}
\end{figure}

\end{document}